\documentclass[dvips]{acta}
\usepackage{supertabular,lscape,epsfig}
\usepackage{amssymb}
\usepackage{amsmath}
\usepackage[export]{adjustbox}
\SetPages{0}{0}

\SetVol{00}{0000}

\begin{document}

\begin{Titlepage}
\Title{Analysis of The Most Precise Light Curves of HAT-P-36 Detrended From Spot Signals}

\Author{S. Yal\c{c}{\i}nkaya$^1$, \"{O}.~Ba\c{s}t\"{u}rk$^1$, F. El Helweh$^{2}$, E. M. Esmer$^1$,  O.~Y\"{o}r\"{u}ko\u{g}lu$^{1}$, M. Y{\i}lmaz$^1$, H.V. \c{S}enavc{\i}$^1$, T. K{\i}l{\i}\c{c}o\u{g}lu$^1$ and S. O. Selam$^1$}
{$^1$Ankara University, Faculty of Science, Astronomy \& Space Sciences Department, Tandogan, TR-06100, Ankara, Turkey\\
$^2$Bilkent University, Science Faculty, Physics Department, TR-06800 Ankara, Turkey\\
e-mail: yalcinkayas@ankara.edu.tr}

\Received{Month Day, Year}
\end{Titlepage}

\Abstract{We study the most precise light curves of the planet-host HAT-P-36 that we obtained from the ground primarily with a brand-new 80 cm telescope (T80) very recently installed at Ankara University Kreiken Observatory (AUKR) of Turkey and also from the space with Transiting Exoplanet Survey Satellite (TESS). The main objective of the study is to analyze the Transit Timing Variations (TTV) observed in the hot-Jupiter type planet HAT-P-36\,b, a strong candidate for orbital decay, based on our own observations as well as that have been acquired by professional and amateur observers since its discovery (Bakos {\it et al.} 2012). HAT-P-36 displays out-of-transit variability as well as light curve anomalies during the transits of its planet due to stellar spots. We collected and detrended all the complete transit light curves we had access to from these anomalies, which we modeled with {\sc EXOFAST} (Eastman {\it et al.} 2013) and measured the mid-transit times forming a homogeneous data set for a TTV analysis. We found an increase in the orbital period of HAT-P-36\,b at a rate of 0.014 seconds per year from the best fitting quadratic function, which is only found in the TTV constructed by making use of the mid-transit times measured from detrended light curves, against an expectation of an orbital decay based on its parameters. We refined the values of these system parameters by modelling the Spectral Energy Distribution of the host star, its archival radial velocity observations from multiple instruments, and most precise transit light curves from the space and the ground covering a wide range of wavelengths with {\sc EXOFASTv2} (Eastman 2017). We also analyzed the out-of-transit variability from TESS observations to search for potential rotational modulations through a frequency analysis. We report a statistically significant periodicity in the TESS light curve at $4.22 \pm 0.02$ days, which might have been caused by instrumental systematics but should be tracked in the future observations of the target.} {stars: individual: HAT-P-36 - methods: observational, transit timing variation - techniques: transit photometry}

\section{INTRODUCTION}
Owing to their large radii and short orbital periods, hundreds of hot-Jupiters have been discovered with the transit method so far. These giant exoplanets in the close vicinity of their stars complicate our understanding of their formation  because it is highly unlikely that they formed where they are observed now, although it was claimed to be possible under special conditions (Bailey \& Batygin 2019). Instead, they are theorized to form far from their parent star, beyond the snow line where the ice can act as glue to help form the core, and then migrate inwards to their known positions (Mordasini {\it et al.} 2015). Due to their short distances from their parent stars, strong tidal forces should have crucial impacts on their orbits. When a planet on an eccentric orbit passes through the periastron, its orbital energy decays with tidal dissipation by the host star leading to its transfer to a tighter and less eccentric orbit. 

In order to investigate the formation and orbital evolution of hot-Jupiters, precise and accurate observational data are mandatory. Detecting the anomalies on the orbital parameters  (i.e. orbital decay or perturbations) relies on the precision and the timespan of the data. Effects like orbital decay may be small but they cumulatively increase within time, making them easier to be detected with precise observations spanning a longer baseline.

HAT-P-36 b is a short period (P $\sim1.33$ days) hot-Jupiter (R$_p$ $\sim1.29$ R$_{\rm jup}$, M$_p$ $\sim1.76$ M$_{\rm jup}$) discovered by Bakos {\it et al.} (2012) around a solar-like star ($T_{\rm eff}$ = 5534 K, this work). An orbital decay is expected based on its physical and orbital parameters (Essick \& Weinberg 2016). The system has been observed many times photometrically to refine ephemerides, investigate Transit Timing Variations (TTVs), and update physical parameters (Wang {\it et al.} 2019, Edwards {\it et al.} 2020, Chakrabarty \& Segupta 2019, Mancini {\it et al.} 2015, Maciejewski {\it et al.} 2013). Mancini {\it et al.} (2015) also found that its orbit is aligned from the Rossiter-McLaughlin Effect (RME) it displays in its radial velocities during the transits of the planet. W\"{o}llert {\it et al.} (2015) and Ngo {\it et al.} (2016) observed the system with an adaptive optics system and found no visual companions. Lillo-Box {\it et al.} (2018) investigated exotrojans in Lagrange points T1 and T4 but they were only able to put mass and radius limits based on their photometric and spectral observations. 

HAT-P-36 is known to display activity-induced, wavelength-dependent light curve modulations due to surface spots, overlapped by the planet disc during some transits (Mancini {\it et al.} 2015). These anomalies cause apparent shifts in the mid-transit times as well as differences in the measurements of the transit depth. This modulation on HAT-P-36 light curves manifests itself in the out-of-transit fluxes as well, providing us a means of determining the rotation period of the host star in better precision, which in turn can enable us better constrain the system's age through tidal-chronology (Gallet 2020). Gaussian Processes (GP), frequently used in modelling correlated noise in time-series data, can be employed to model and then remove spot-induced asymmetries on transit light curves so that mid-transit times and transit depths can be measured with better precision and accuracy from them. By combining these "clean" light curves with radial velocity observations, absolute parameters can be obtained with the help of theoretical stellar models. Semi-empirical radius of the host star, which can be measured by modelling its Spectral Energy Distribution (SED) based on broad-band photometry and its distance, helps constrain the absolute parameters even further. 

We observed the target several times with the brand new 80 cm telescope T80 located at Ankara University Kreiken Observatory (AUKR) and 1 m telescope T100 in T\"UB{\.I}TAK National Observatory of Turkey (TUG). We collected all the complete and precise light curves from amateur and professional observers. We obtained light curves of the system from the Transiting Exoplanet Survey Satellite (TESS), observed during the Sector-22 in the short-cadence mode (2 minutes), and formed the largest and most precise set of light curves for the system. We then detrended these light curves from linear effects as well as spot-induced modulations and systematics due to instrumental effects with GP. Observations and data processing are described in detail in Section-2. We performed a global modelling of the system by making use of the most precise light curves covering a wide range of wavelengths, archival radial velocities from multiple telescopes (Bakos {\it et al.} 2012, Mancini {\it et al.} 2015, Lillo-Box {\it et al.} 2018), SED of the host star from its broad-band photometry and also the atmospheric parameters obtained from high resolution spectroscopy in previous analyses (Bakos {\it et al.} 2012, Mancini {\it et al.} 2015 (Section-3.2). Finally, we analyzed the Transit Timing Variations (TTV) observed in the HAT-P-36 system to investigate a potential period decrease based on this longest baseline of observations in time, ever analyzed for this particular system. In section 3, we describe the global modelling as a result of which we obtain planetary and stellar parameters as well as our TTV analysis (Section-4). We present a discussion of our results in Section-5.

\section{OBSERVATIONS AND DATA REDUCTION }
\label{sec:observations}
We observed 6 transit events of HAT-P-36\,b with the recently installed \emph{"Prof. Dr. Berahitdin Albayrak Telescope"} (T80) at AUKR in Turkey. The telescope has an 80 cm diameter primary mirror with f/7 focal ratio, which translates into $37^{\prime\prime}$/mm plate scale. With the focal reducer of $0.69 \times$, the plate scale is reduced to $53.4 ^{\prime\prime}$/mm and $11^{\prime}.84 \times 11^{\prime}.84$ field of view (FoV) is achieved on a $1024 \times 1024$ back-illuminated CCD with a pixel size of 13 $\mu$m. We used Sloan-r$^{\prime}$ filter for all six observations. We observed a transit of HAT-P-36\,b with the T100 telescope on 22 Feb 2021 at TUG, which has a 1 m diameter primary mirror with F/10 focal ratio  ($21^{\prime\prime}$/mm plate scale), connected to a back-illuminated $4096 \times 4096$ CCD, giving an effective FoV of $21^{\prime} \times 21^{\prime}$. The readout is completed in $\sim45$ seconds, hence we made use of $2 \times 2$ binning mode to reduce it to $\sim15$ seconds. We also used an auto-guider system in the T100 observations so the change of the pixel coordinates of the target were no more than a few pixels throughout the night. On both telescopes, we used the defocusing photometry technique to reduce photon noise and mitigate the effects of minor tracking problems and flat fielding (Southworth {\it et al.} 2009, Ba\c{s}t\"{u}rk {\it et al.} 2015). We provide a log of our observations in Table-1. 

\begin{table}
	\centering
	\caption{A log of photometric observations analyzed within this study for the first time. Nightly average of the photometric measurement uncertainties ($\sigma_{ph}$), Photon Noise Rate (PNR), and the $\beta$-factors quantifying the white and red-noise, respectively are given in 6-8 columns.}
	\label{tab:ph_observations}
	\begin{tabular}{ccccccccc} 
		\hline
		Obs. & Starting & Facility & Exp.Time & Filter & $\sigma_{ph.}$ & PNR & $\beta$\\
                Number &Date [UT] & & [s] &  & [ppt]  & [ppt] &  \\ 
		\hline
		1 & 2021-02-06 & AUKR-T80 & 130 & Sloan-$r^{\prime}$ & 2.17  & 1.35 & 0.85 \\
		2 & 2021-02-22 & AUKR-T80 & 130 & Sloan-$r^{\prime}$ & 1.68  & 1.07 & 0.89 \\
		3$^{a}$ & 2021-02-22 & TUG-T100 & 130 & Bessel-$R$ & 1.01 & 0.87 & 0.88 \\
		4 & 2021-02-26 & AUKR-T80 & 60 & Sloan-$r^{\prime}$ & 2.08  & 2.35 & 0.45 \\
        5$^{a}$ & 2021-05-06 & AUKR-T80 & 130 & Sloan-$r^{\prime}$ & 1.17 & 0.99 & 1.04 \\
        6$^{a}$ & 2021-05-10 & AUKR-T80 & 130 & Sloan-$r^{\prime}$ & 1.02 & 1.04 & 0.87 \\
        7$^{a}$ & 2021-05-14 & AUKR-T80 & 130 & Sloan-$r^{\prime}$ & 0.62 & 0.70 & 0.77 \\
        8$^{a}$$^{b}$ & 2020-02-18 & TESS & 120 & TESS & 0.7 & 0.28 & 0.91 \\
		\hline
	\end{tabular}
\label{tab1}
    \centering
    {$^a$used in global modelling\\$^b$parameters are for the binned light curve used in global modelling}
\end{table}

 We used {\sc AstroImageJ} (Collins {\it et al.} 2017) to perform data reduction and photometry on the images acquired in all of our observations in the standard manner. We made use of an ensemble of comparison stars, potential variability of each of which we investigated. During these investigations, we noticed the variability in TYC 3020-2195-1 star, which is also noticed by American Association of Variable Star Observers, AAVSO\footnote{https://www.aavso.org/vsx/index.php?view=search.top}, $\sim3'$ away from the target. We then extracted light curves with respect to the best set of comparison stars in terms of proximity, magnitude, color, and stability. Finally, we detrended the light curves for the airmass-effect by making use of the linear trend in the out-of-transit data.  

We obtained additional light curves from TESS observations, literature and amateur observers from Exoplanet Transit Database (ETD\footnote{http://var2.astro.cz/ETD/}). We collected the light curves from the literature and ETD, only if they cover the full transits. Observers reporting to ETD provide an integer value (from 0 to 5) for data quality. We limited this criterion to a  data quality value of 3 while selecting light curves from the ETD. We converted every time frame of observations to barycentric dynamical time (BJD-TDB) and also calculated the airmass with a Python script, then airmass-detrended the light curves in same manner as our observations.

TESS observed 18 transit events of HAT-P-36\,b during the Sector-22 of the mission and the star has been chosen as an object of interest (TOI 1810.01) with TIC ID 373693175. Therefore it obtained 2 minute-cadence observations, for which data validation files were available. We used the fluxes listed in the LC\_DETREND column, which are extracted and detrended by the TESS Science Processing Operations Center (SPOC) pipeline (Jenkins {\it et al.} 2016). Then we modelled selected light curves (7 from our observations, 19 from literature, 46 from ETD and 18 from TESS) with the browser version of {\sc exofast}\footnote{https://exoplanetarchive.ipac.caltech.edu/cgi-bin/ExoFAST/nph-exofast} (Eastman {\it et al.} 2013) to calculate mid-transit times, Photon Noise Rate (PNR, Fulton {\it et al.} 2011) and red noise parameter $\beta$ (Winn {\it et al.} 2008) as described in Ba\c{s}t\"{u}rk {\it et al.} (2020).

We provide several noise statistics to quantify the quality of our photometric observations in Table-1. These quality parameters for AUKR T80 varies from each other due to different weather conditions at the dates of observation. Although AUKR observations are affected relatively more from light pollution compared to TUG observations, high-quality optical system of T80 telescope makes it possible to have precise observations with comparable photometric quality and noise statistics to T100.

Our observations are available in the machine readable format and can be found in the CDS linked to this study. We provide detrended and normalized light curves within the online data; however, we can also provide raw data sets on request from the corresponding author.

\subsection{Detrending Light Curves with Gaussian Processes}
Activity-induced stellar spots are known to cause hump-like structures or at least asymmetries on transit profiles in the light curves of active transiting planet hosts. The profile center, the ingress, and the egress are all affected by the spot-crossing events in the transit chord, hence the measurements of the mid-transit times, which rely heavily on their timings, become ambiguous. This poses a major problem in the detection of potential variations in the orbital period of the target by analyzing the variations in these timings. There are additional red (correlated) noise sources such as pixel response variations, changes in the positions of the target and comparison stars on the CCD and focus, occasional clouds etc., which worsen the problem. Therefore we needed to treat the spot-induced signals in the transit light curves of HAT-P-36\,b as correlated (red) noise and detrend them from the disruptive effects of the spots together with other red noise sources. This approach would also help us to determine the depths of transits in better precision as well, which in turn improves the precision of the parameters depending on the measurements of them in the light curves we used for global modelling, provided that the white noise level in the data is preserved and they are not oversmoothed.

\begin{figure}[h]
\begin{center}
\includegraphics[width=1\textwidth]{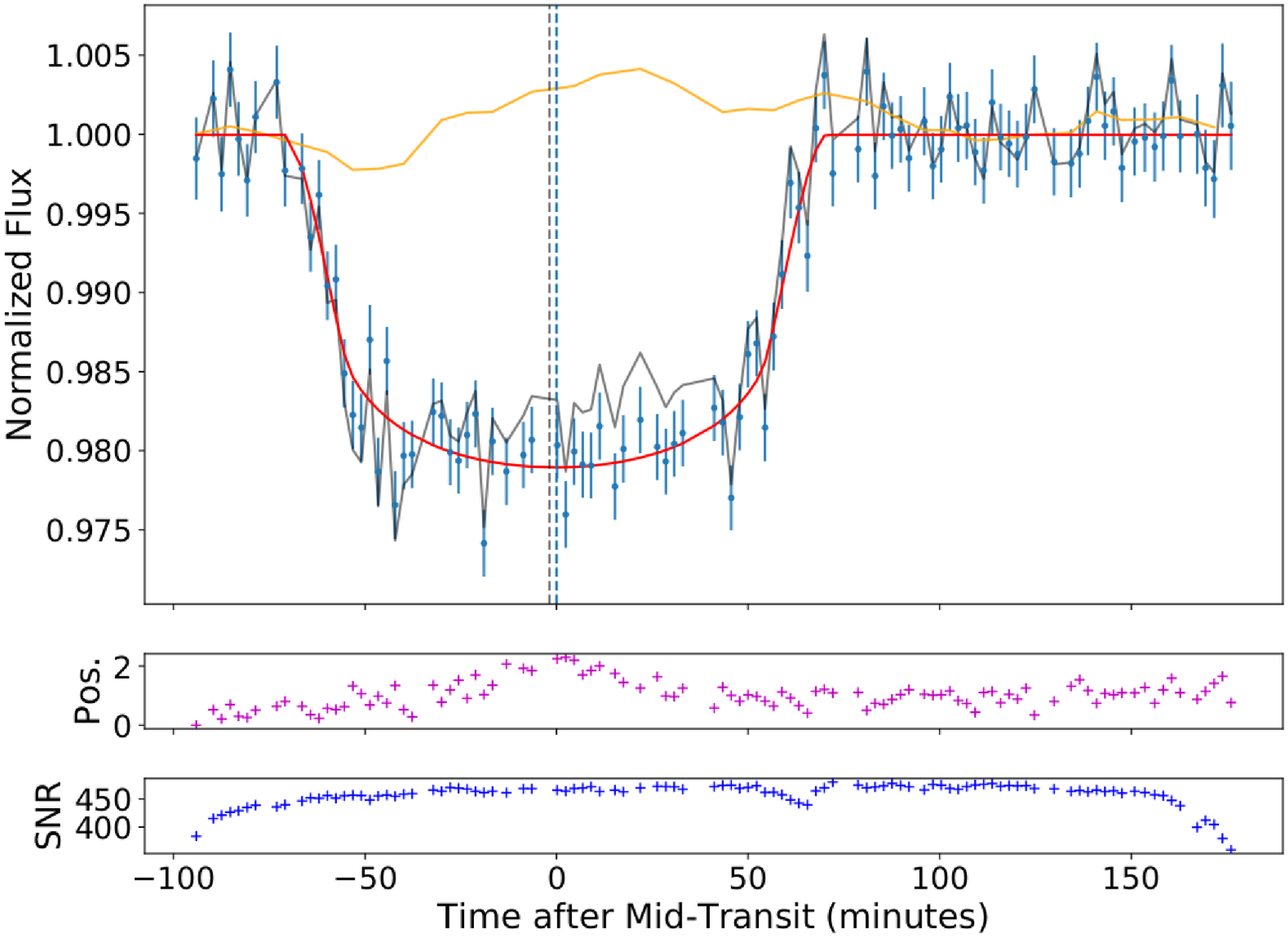}
\setlength{\abovecaptionskip}{15pt}
\FigCap{T80 transit light curve (blue data points) of HAT-P-36 b acquired on 6 February 2021 and detrended with the stochastic model (orange) from the correlated noise (upper panel). Dark gray curve shows the undetrended light curve while the {\sc exofast} transit model for the detrended light curve is given with the red continuous curve. There is a $\sim110$ second-difference between the vertical blue and gray dashed lines showing the mid-transit times derived from the {\sc exofast} models of detrended and raw light curves, respectively. Middle panel shows the change in the distance of the position of the target to its initial position on the CCD in pixels.  The bottom panel illustrates the variation in the SNR for the target.}
\end{center}
\end{figure}

In order to detrend all the light curves of HAT-P-36\,b we used in this study, we made use of a quasi-periodic kernel for the stochastic part of a light curve model ($C_{i,j}$) in the form the Equation-1 in addition to the deterministic part of the model, which is essentially the photometric model of a transit ($\mathcal{M}_{\rm tr}$).

\begin{equation}{
    C_{i,j} = \frac{B}{2 + C} e^{-\frac{|t_i - t_j|}{L}} [cos(\frac{2\pi |t_i - t_j|}{P}) + (1 + C)]
    }
\end{equation}

where $C_{i,j}$ is an arbitrary element of the matrix approximating the quasi-periodic kernel, $B$ and $C$ are parameters controlling the amplitude, $L$ controls the length-scale, and P is the candidate period of the quasi-periodic function. This part of the Gaussian Processes (GP) fitting applied to detrend the light curves forms the stochastic part ($C_{i,j}$) that fits the red-noise.

The photometric model ($\mathcal{M}_{\rm tr}$) (Equation-2), on the other hand, is composed of a dilution factor ($D$), mean of the out-of-transit flux ($M$), and the transit model for a given instrument $\mathcal{T}(t)$, which is defined by the transit parameters and by the instrument-dependant quadratic limb-darkening coefficients $q_1$ and $q_2$. 

\begin{equation}{
\mathcal{M}_{\rm tr}(t) = [\mathcal{T}(t) \times D + (1-D)] (\frac{1}{1+D} \times M)
}
\end{equation}

 We used the {\sc juliet} code (Espinoza {\it et al.} 2019) in Python to fit the light curves as described.  We assigned normal priors to transit parameters for the deterministic part, centers of which are set to the values from Wang {\it et al.} (2019) with 1-standard deviations equal to that given in the same study. The transit parameters to fit were radius ratio (R$_{p}$ / R$_{\star}$), impact parameter (b), semi-major axis scaled to stellar radius (a / R$_{\star}$), orbital period (P$_{\rm orb}$), and the mid-transit time ($T_c$). We assigned normal priors to instrument-related parameters (dilution factor (D), mean out-of-transit flux (M)) accordingly with light curve parameters, while the parameters of the stochastic part of the fit were assigned to uninformative, log-uniform priors. The combined GP model to detrend the light curves with the form $\mathcal{M} = \mathcal{M_{\rm tr}} + \epsilon(\rm t)$ was obtained from {\sc juliet}. We evaluated the combined model with stochastic and deterministic parts for all light curve points and subtracted the stochastic part from the data at the end to have the detrended transit light curve. We provide an example light curve (blue) with the base transit model (black), detrended from the correlated noise with the help of the stochastic model (orange) in Figure-1 for T80 observation on 6 Feb 2021, when a spot-like signal was observed after the ingress. The general upward trend in the light curve until mid-transit was most probably caused by the changes in pixel position of the target (middle panel of Fig.1), which in turn seemed to affect its SNR (bottom panel of Fig.1). The depression closer to the ingress can be of stellar origin since its amplitude is in agreement with the out-of-transit amplitude noticed in HATNet data by Mancinin et al. (2015) and in TESS data by us.

\subsection{Light Curve Selection for Global Modelling}
We modelled all 95 detrended light curves with the same version of {\sc exofast}, as we did for all of the light curves before detrending. We calculated $\beta$ and PNR parameters again and noticed that the red noise parameter $\beta$ was reduced (approached to 1) significantly. We then selected best light curves in each passband and also the most precise light curves with the least amount of correlated noise, especially during the ingress/egress times, before detrending. We phase-folded the TESS light curves by making use of the orbital period from the Data Validation Time Series (dvt file), binned the data to have a data point for every 2 minutes, and then switched back to the BJD-TDB time frame by using the same period again. As a result, we obtained a light curve with a higher signal-to-noise Ratio (SNR), which also reduced the integration time for global modelling dramatically. Light curves we used in global modelling and TTV analysis are given in Figure-2 while those we used only in the TTV analysis from our observations are provided in Figure-3

\begin{figure}[h]
\includegraphics[width=1\textwidth]{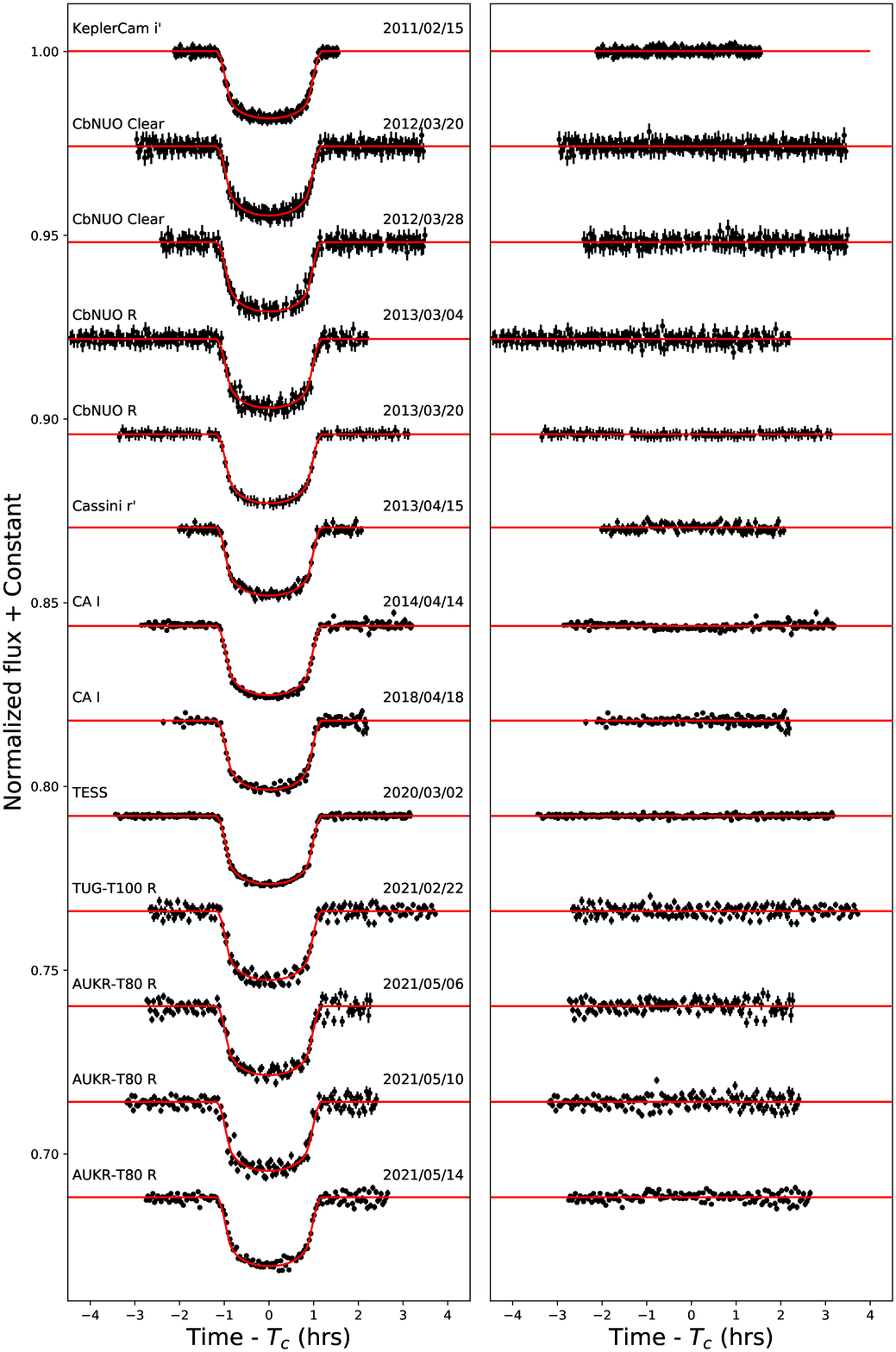}
\vspace*{-25mm}
\FigCap{Individual transit light curves, their {\sc EXOFASTv2} models (left panel), and their residuals (right panel). Data points and their errorbars are in black, while their models are illustrated with red continuous curves. KeplerCam observation is from Bakos {\it et al.} (2012), CbNUO observations from Wang {\it et al.} (2019), CA and Cassini observations from Mancini {\it et al.} (2015).}
\end{figure}

\clearpage
\begin{figure}
    \begin{center}
    \includegraphics[width=1\textwidth]{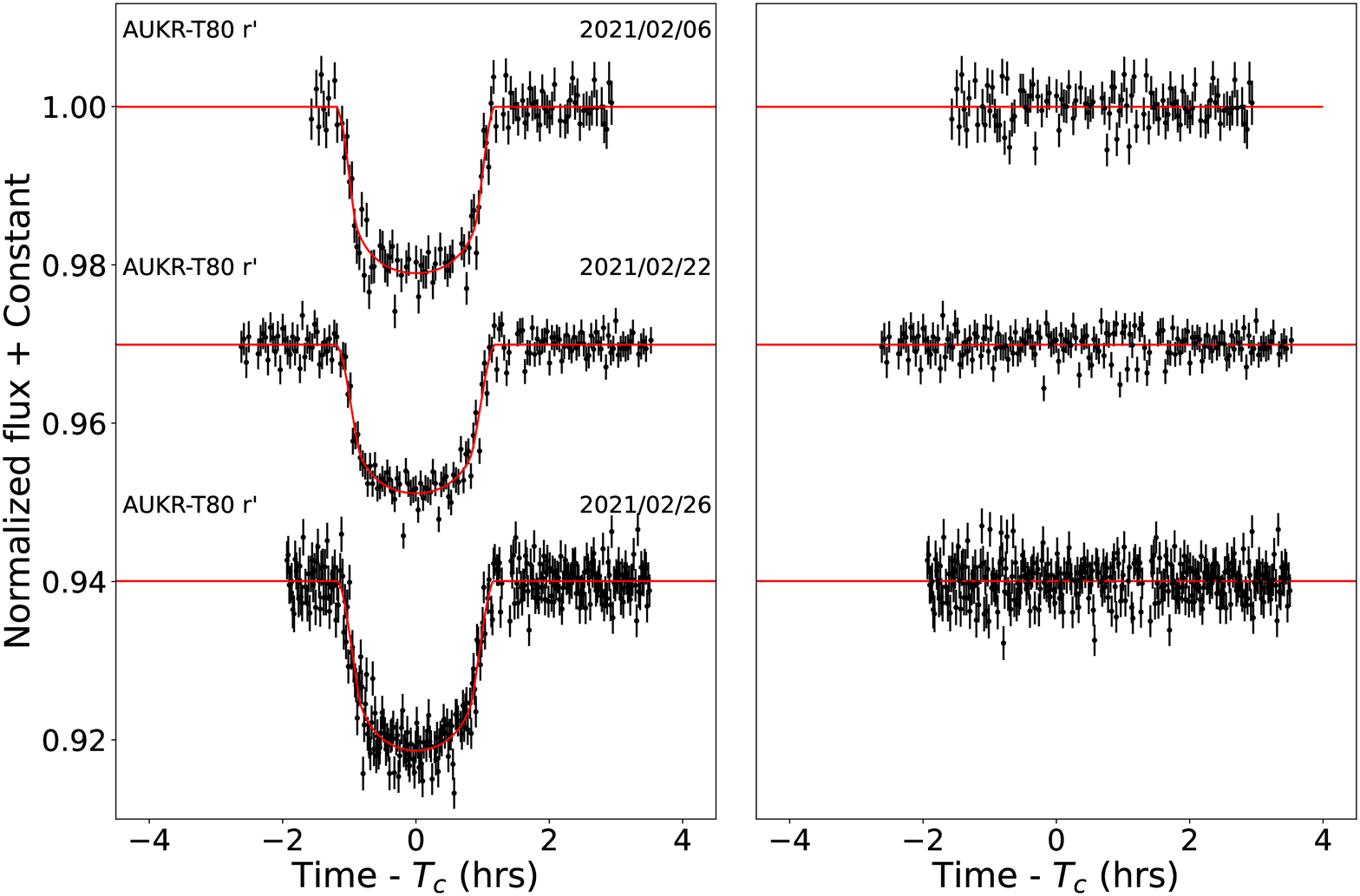}
    \vspace*{4mm}
    \caption{T80 Light curves that we didn't use in global modelling but used in TTV analysis shown in left panel with black dots. {\sc EXOFAST} model to determine transit mid points is shown with red continuous line. Residuals are shown in right panel.}
    \end{center}
\end{figure}

\section{ANALYSIS AND RESULTS }
\subsection{Stellar Parameters}
We fitted the SED of the star using Experiments in Stellar Astrophysics (MESA) Isochrones \& Stellar Tracks (MIST) bolometric correction grid\footnote{$http://waps.cfa.harvard.edu/MIST/model\_grids.html$} (Choi {\it et al.} 2016) with {\sc EXOFASTv2} based on broadband photometry from different passbands used mostly in space-borne observations (listed in Table-2). We also provided parallax measurements from Gaia DR2 as a Gaussian prior after the addition of an offset value ($0.082^{\prime\prime}$) noticed by Stassun \& Torres (2018), propagating the offset to the uncertainty in parallax as $0.033^{\prime\prime}$. We allowed the V-band extinction value (A$_{\rm V}$) to vary as a free parameter, but limited it to the line of sight value given by Schlegel {\it et al.} (1998). We noticed that A$_{\rm V}$ is small, which is why distance of the star theoretically calculated by Bakos {\it et al.} (2012) is in agreement with that from the Gaia measurements. We adopted stellar metallicity [Fe/H] and surface gravity $log~{\rm g}$ parameters from Bakos {\it et al.} (2012) which are confirmed by Mancini {\it et al.} (2015). In general, the $T_{\rm eff}$ value from the SED fitting is less precise than that is derived from the spectral analysis. However the $T_{\rm eff}$ value that we determined from our SED analysis is exactly equal to the mean value of those from previous analyses (Mancini {\it et al.} 2015, Bakos {\it et al.} 2012). Nevertheless, we preferred to use $T_{\rm eff}$ value from our analysis in global modelling ($5590\pm120$ K). We also provided the stellar radius (R$_{\star}$) value, derived from SED analysis as the center of a normal prior ($1.009\pm0.037$ $R_\odot$) during global modelling. Minimum values of uncertainties for $T_{\rm eff}$ and R$_{\star}$ are constrained by default according to Tayar {\it et al.} (2020). Our results are shown in Table-3.

\begin{figure}
\begin{center}
\includegraphics[width=1\textwidth]{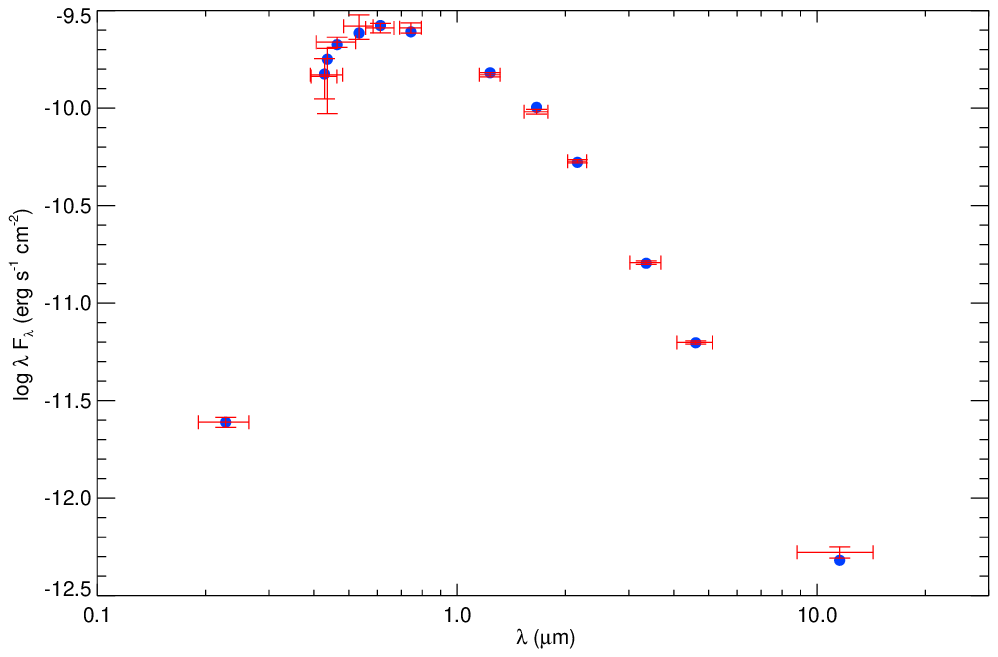}
\FigCap{Fluxes in broadband filters of HAT-P-36 (red data points with error bars) and corresponding fluxes from the best fitting model (blue dots).}
\end{center}
\end{figure}

\begin{table}
\centering
\caption{Passband Brightnesses of HAT-P-36.}
\label{tab2}
\begin{tabular}{ccc}  
\hline
\hline
Passband & $\lambda_{\rm eff} $ (\AA) & Magnitude \\
\hline
\multicolumn{3}{l}{APASS-DR9 (Henden {\it et al.} 2016)}\\
\hline
Johnson B & 4378.1 & $13.174\pm0.4$ \\
SDSS g' & 4640.4 & $12.58\pm0.065$ \\
SDSS r' & 6122.3 & $12.0.97\pm0.061$ \\
SDSS i' & 7439.5 & $11.883\pm0.065$ \\
\hline
\multicolumn{3}{l}{GALEX (Bianchi {\it et al.} 2017)}\\
\hline
galNUV & 2274.4 & $18.2258\pm0.0645$ \\
\hline
\multicolumn{3}{l}{2MASS (Cutri {\it et al.} 2003)}\\
\hline
$J_{2MASS}$ & 12350.0 & $11.046\pm0.027$ \\
$H_{2MASS}$ & 16620.0 & $10.723\pm0.030$ \\
$K_{2MASS}$ & 21590.0 & $10.603\pm0.021$ \\
\hline
\multicolumn{3}{l}{All WISE (Cutri {\it et al.} 2021)}\\
\hline
WISE1 & 33526.0 & $10.594\pm0.023$ \\
WISE2 & 46028.0 & $10.633\pm0.020$ \\
WISE3 & 115608.0 & $10.489\pm0.072$ \\
\hline
\multicolumn{3}{l}{Tycho-2 (H{\o}g {\it et al.} 2000)}\\
\hline
BT & 4280.0 & $13.168\pm0.253$ \\
VT & 5340.0 & $12.238\pm0.156$ \\
\hline
\end{tabular}
\end{table}

HAT-P-36\,b is a moderately active star with logR'$_{HK}$ = $-4.636\pm0.066$ dex (Mancini {\it et al.} 2015). The magnetic activity-induced stellar spots are even noticed to be affecting the transit profiles in the cases of spot-crossing events in the transit chord in both space and ground-based observations. In a longer timescale, surface brightness inhomogeneities modulate the out-of-transit light curve too. We investigated if such a modulation was observed by TESS during the Sector-22 of observations in the Pre-search Data Conditioning Simple Aperture Photometry (PDCSAP) fluxes derived by the SPOC pipeline.

We simply cut the transits on the entire light curve for Sector-22 to obtain the out-of-transit light curve. We then removed some data points at the beginning and end of two segments separated by the gap at the data downlink time due to their insufficient precision and potentially misleading accuracy. We then performed a frequency analysis with a Lomb-Scargle periodogram (Lomb 1976, Scargle 1982) by using {\sc astropy} package (Astropy Collaboration {\it et al.} 2013, 2018). We found a statistically significant peak at 0.2373 day$^{-1}$ with a False Alarm Probability of practically zero thanks to very high SNR. We checked the SAP (Simple Aperture Photometry) fluxes from the SPOC pipeline for the star TYC 3020-2195-1, which is in close vicinity of HAT-P-36, which has a similar brightness (m$_{\rm V}$ = $11^{\rm m}.62$) and color (B - V = $0^{\rm m}.11$). We did not find a similar frequency in its light curve except for a very strong peak at 14.2526 day$^{-1}$, which probably corresponds to the period of the intrinsic variability of this star (${\rm P} \sim 101.088$ minutes), which was already found to be variable by AAVSO observers. The periodogram of this star is very rich, deserving a careful look to study its variability in particular, which is however, out of the scope of this work. 

Since the amplitude and probably of the frequency of the out-of-transit variation of TESS PDCSAP lightcurve of HAT-P-36 is observed to be changing before and after the data downlink gap, we treated these two segments of the PDCSAP light curve separately and repeated the frequency analysis for each of them. Although the FAP (False Alarm Probability) of the signal in the first segment is found to be significant, its frequency is significantly smaller (0.1950 day$^{-1}$) whereas the frequency derived from the second segment (0.2267 day$^{-1}$) is in strong agreement with that from the entire Sector-22 PDCSAP light curve. We then questioned if these periodicities are observed in the spot-crossing events in the transit profiles. However, we were not able to find any periodicity in the change of the positions of the spot-induced asymmetries on transit profiles. We then fitted the entire light curve with a perfect sinusoidal, initial parameters of which were set to the values derived from its Lomb-Scargle periodogram with the Levenberg-Marquardt algorithm and obtained the period of the variation as $4.22 \pm 0.02$ days. Although this value is close to the momentum dumps appearing at $\sim$5 days, the variation was continuous. In addition, the length of a single TESS sector is not adequate to probe the rotation period found by Mancini {\it et al.} (2015) from the complete HAT-Net data set as $15.3 \pm 0.4$ days because it is close to and slightly longer than the orbital period of the spacecraft (13.7 days) causing background variations due to the phase dependency of the reflected sunlight.When we restrict the power spectra of its SAP flux light curves for both HAT-P-36 and this nearby star to lower frequencies we obtain very similar frequency at 0.066 days$^{-1}$ (15.01 days). Therefore, although a similar frequency is found with strong spectral leakage in PDCSAP fluxes of HAT-P-36 too, the power spectrum might have been affected by the systematics of the instruments.

As a result we report the periodicity we found from the out-of-transit variability in the PDCSAP-fluxes as 4.22 days for future observations of the target, especially TESS observations during the sector-49 from Feb-26 to Mar-26 in 2022, to verify. Mancini {\it et al.} (2015) investigated a rotation period of 4.57 days in their study assuming that the signal of the same spot had been observed in the transit chord, separated by four days. They found it implausible considering the sky-projected rotation rate (v sini$_{\star}$) derived from spectroscopy by Bakos {\it et al.} (2012) as 3.58 km/s. As a result, we adopt the rotation period value found by Mancini {\it et al.} (2015) as $15.3 \pm 0.4$ days from HAT-Net data in our estimation of the age of the host-star from tidal-chronology.

\subsection{Global Modelling}
We modelled the detrended transit light curves that we selected and the radial velocity (RV) data together with the information from the stellar evolution models for HAT-P-36 simultaneously. We provided $T_{\rm eff}$ and $R_{\star}$ parameters from our SED-fitting results, [Fe/H] from Bakos {\it et al.} (2012) as the central values of Gaussian priors. Note that $log~{\rm g}$ was not used in global modelling because the mean stellar density ($\rho_\star$) can be employed in the stellar evolution models instead, which can be constrained within better precision from transit photometry by making use of the Kepler's third law. It has also the potential to be more accurate than the $log~{\rm g}$ value derived from high resolution spectroscopy because it is degenerate with other factors controlling the spectral line profile. EXOFASTv2 interpolates for the quadratic limb darkening parameters in the tables provided by Claret (2017) for the TESS band and Claret \& Bloemen (2011) for the other bands based on the atmospheric parameters of the host star and uses it as a Gaussian prior. Eastman {\it et al.} (2019) recommends the selection of the passband with a similar transmission curve to an unsupported passband when it is the case. Therefore we employed CoRoT passband for clear observations (Wang {\it et al.} 2019), and Johnson-R passband for Bessel-R, which is the passband used in T100 observation. All other parameters were adjusted by sampling from uniform distributions, initial values of which are set to the values derived from preliminary analysis in order to reduce the integration time. {\sc EXOFASTv2} can model (RME) effect but we discarded the RV data during the transit since its modelling wouldn't have any impact on the absolute parameters. In total, we used 12 RV points from TRESS (Bakos {\it et al.} 2012), 11 from HARPS-North (4 from Mancini {\it et al.} (2015) and 7 from Lillo-Box {\it et al.} (2018) ) and 7 from CARMENES (Lillo-Box {\it et al.} 2018) after converting the timings to BJD-TDB format. {\sc EXOFASTv2} automatically fits the velocity offset (i.e. $V_\gamma$) for each data sets from different telescopes. Unfortunately, secondary eclipse (i.e. occultation) of HAT-P-36\,b hasn't been observed yet and it can not be recovered in the TESS observations as well by phase-folding and binning. We calculated the predicted occultation depth as 203 ppm with using guide from Wong {\it et al.} (2021) but the standart deviation of residuals of 2 minutes binned TESS data is 400 ppm. This makes it challenging to put a tight constraint on the orbital eccentricity from the RV data alone and to determine whether the orbit is circular or eccentric as a result. Therefore we made two global models, one with the assumption of a circular orbit and the other with the eccentricity value derived from RV curve. We made a comparison of the BIC (Bayesian Information Criteria) values from the two models, which favors the circular orbit with $\Delta$BIC = 6, therefore we adopted the results from the model based on the circular orbit assumption.

We used MIST to determine the age and the mass of the star. The $\rho_\star$ derived from the transit photometry and the prior on the $R_{\star}$ value affect the mass value. During the main sequence lifetimes, these absolute parameters of low mass stars, such as HAT-P-36, change very slowly and in smaller amounts than the theoretical models can precisely predict, which is why age determination from isochrone models are usually not accurate although they have small error bars that seem reasonable. We found HAT-P-36 is $\sim9$ Gyrs old from the MIST models, which makes it older than expected from its moderate magnetic activity and reported stellar rotation rate (Mancini {\it et al.} 2015). In most cases isochrone age is overestimated for the same reason. On the other hand, gyrochronological age could be underestimated due to the fact that the angular momentum transfer from the fast orbiting planet to the star during the evolution of the system is not accounted for. Therefore we made use of the {\sc tatoo}\footnote{https://github.com/GalletFlorian/TATOO/} code (Gallet 2020) which takes into account the angular momentum transfer and hence calculates a "tidal-chronology" age. {\sc tatoo} code simply estimates the angular momentum transfer from the planet dissipated by the host star and calculates a corrected gyrochronological age. Input parameters are orbital period, masses of the planet and the star, which we set as we found from our global modelling, and the stellar rotation period, which we adopted from Mancini {\it et al.} (2015). {\sc tatoo} code also calculates an uncorrected gyrochronological age which is much less than the tidal-gyrochronological age as expected. We didn't adopt the periodicity we found from the out-of-transit variability (4.22 days) in TESS observations, which also makes the the tidal-chronological age unrealistic and used the rotational period found by Mancini {\it et al.} (2015) instead. We calculated the gyrochronological as as $1.52 \pm 0.04$ Gyrs, however, tidal interactions with the host star may have it spun-up to look younger than it actually is. When this is accounted for through tidal-chronology, we found the age to be $3.65 \pm 0.33$ Gyrs. The results of all three methods for estimation of the stellar age are listed in Table-3.

The parameter values from our global model are also provided in Table-3, on which the light curve models given in Figure-2 and RV models in Figure-5, SED models in Figure-4 are based. 

\begin{figure}[h]
\begin{center}
\includegraphics[width=\textwidth]{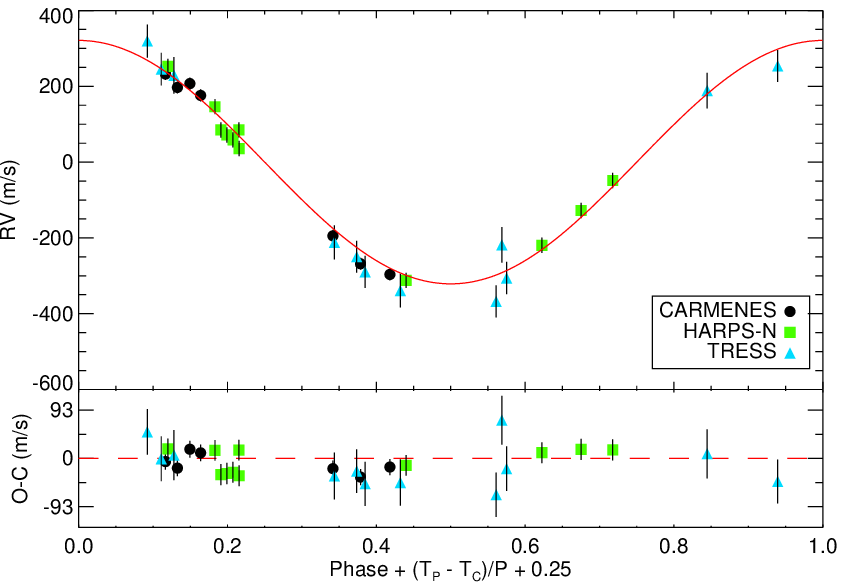}
\FigCap{Radial velocity observations from CARMENES (black data points), HARPS-N (green data points) and TRESS (teal data points). The red curve represents our best-fitting Keplerian model. Residuals from the model are given in the bottom panel.}
\end{center}
\end{figure}

\providecommand{\bjdtdb}{\ensuremath{\rm {BJD_{TDB}}}}
\providecommand{\feh}{\ensuremath{\left[{\rm Fe}/{\rm H}\right]}}
\providecommand{\teff}{\ensuremath{T_{\rm eff}}}
\providecommand{\teq}{\ensuremath{T_{\rm eq}}}
\providecommand{\ecosw}{\ensuremath{e\cos{\omega_*}}}
\providecommand{\esinw}{\ensuremath{e\sin{\omega_*}}}
\providecommand{\mj}{\ensuremath{\,M_{\rm J}}}
\providecommand{\rj}{\ensuremath{\,R_{\rm J}}}
\providecommand{\me}{\ensuremath{\,M_{\rm E}}}
\providecommand{\re}{\ensuremath{\,R_{\rm E}}}
\providecommand{\fave}{\langle F \rangle}
\providecommand{\fluxcgs}{10$^9$ erg s$^{-1}$ cm$^{-2}$}

\begin{table}
\fontsize{9}{9}\selectfont
\caption{Stellar and Planetary parameters of the  HAT-P-36 system\label{tab:global_parameters}}
\setlength{\tabcolsep}{6pt}
\renewcommand{\arraystretch}{1.2}
\begin{tabular}{llc}
\hline \hline
Symbol & Parameter (Unit) & Value\\
\hline
\multicolumn{3}{l}{Stellar Parameters:} \\
\hline
$M_{\star}$&Mass ($M_\odot$)&$0.961^{+0.055}_{-0.042}$\\
$R_{\star}$&Radius ($R_\odot$)&$1.052^{+0.021}_{-0.018}$\\
$R_{\star,SED}$&Radius from SED ($R_\odot$)&$1.009^{+0.037}_{-0.037}$\\
$L_{\star}$&Luminosity ($L_\odot$)&$0.94^{+0.08}_{-0.07}$\\
$\rho_{\star}$&Density (cgs)&$1.164^{+0.035}_{-0.032}$\\
$\log{g}$&Surface gravity (cgs)&$4.377^{+0.012}_{-0.011}$\\
$T_{\rm eff}$&Effective Temperature (K)&$5534^{+84}_{-83}$\\
$T_{\rm effSED}$&Effective Temperature from SED (K)&$5590^{+120}_{-120}$\\
$[{\rm Fe/H}]$&Metallicity (dex)&$0.249\pm0.094$\\
$[{\rm Fe/H}]_{0}$&Initial Metallicity &$0.265^{+0.085}_{-0.086}$\\
$Age_{iso}$&Isochrone Age (Gyr)&$9.5^{+2.7}_{-3.0}$\\
$Age_{tidal-chro}$&Tidal-Chronology Age (Gyr)&$3.65\pm0.34$\\
$Age_{gyro}$&Gyrochrolology Age (Gyr)&$1.52\pm0.04$\\
$EEP$&Equal Evolutionary Point &$402^{+12}_{-21}$\\
$A_V$&V-band extinction (mag)&$0.021^{+0.013}_{-0.014}$\\
$\varpi$&Parallax (mas)&$3.446\pm0.058$\\
$d$&Distance (pc)&$290\pm5$\\
\hline
\multicolumn{3}{l}{Planetary Parameters:}\\
\hline
$P$&Period (days)&$1.327346702 \pm 0.000000043$\\
$R_{\rm p}$&Radius (\rj)&$1.288^{+0.027}_{-0.024}$\\
$M_{\rm p}$&Mass (\mj)&$1.759^{+0.079}_{-0.069}$\\
$a$&Semi-major axis (au)&$0.02334^{+0.00044}_{-0.00034}$\\
$i$&Inclination (Degrees)&$85.89^{+0.39}_{-0.34}$\\
$T_{eq}$&Equilibrium temperature (K)&$1791\pm29$\\
$K$&RV semi-amplitude (m/s)&$332.7^{+8.2}_{-9.0}$\\
$\rho_{\rm p}$&Density (cgs)&$1.021^{+0.050}_{-0.047}$\\
$logg_{\rm p}$&Surface gravity &$3.42\pm0.02$\\
$\Theta$&Safronov Number &$0.0662\pm0.0021$\\
$\fave$&Incident Flux (\fluxcgs)&$2.34^{+0.16}_{-0.15}$\\
\hline
\multicolumn{3}{l}{Transit Parameters:}\\
\hline
$b$&Transit impact parameter &$0.342^{+0.025}_{-0.030}$\\
$\delta$&Transit depth (fraction)&$0.01580^{+0.00011}_{-0.00012}$\\
$a/R_{\star}$&Semi-major axis in stellar radii &$4.772^{+0.047}_{-0.044}$\\
$\tau$&Ingress/egress transit duration (days)&$0.01213\pm0.00028$\\
$T_{14}$&Total transit duration (days)&$0.09604\pm0.00027$\\
\hline
\multicolumn{3}{l}{Wavelength Parameters:}\\
\hline
$u_{1,Clear}$&linear limb-darkening coeff in Clear &$0.451\pm0.028$\\
$u_{2,Clear}$&quadratic limb-darkening coeff in Clear &$0.212\pm0.034$\\
$u_{1,I}$&linear limb-darkening coeff in I &$0.403\pm0.024$\\
$u_{2,I}$&quadratic limb-darkening coeff in I &$0.322^{+0.031}_{-0.032}$\\
$u_{1,R}$&linear limb-darkening coeff in R &$0.425^{+0.023}_{-0.022}$\\
$u_{2,R}$&quadratic limb-darkening coeff in R &$0.244\pm0.026$\\
$u_{1,i'}$&linear limb-darkening coeff in i' &$0.365\pm0.031$\\
$u_{2,i'}$&quadratic limb-darkening coeff in i' &$0.239^{+0.043}_{-0.042}$\\
$u_{1,r'}$&linear limb-darkening coeff in r' &$0.417\pm0.036$\\
$u_{2,r'}$&quadratic limb-darkening coeff in r' &$0.212^{+0.046}_{-0.045}$\\
$u_{1,TESS}$&linear limb-darkening coeff in TESS Band &$0.392\pm0.028$\\
$u_{2,TESS}$&quadratic limb-darkening coeff in TESS Band &$0.257\pm0.040$\\
\hline
\multicolumn{3}{l}{Auxiliary RV Parameters:}\\
\hline
$\gamma_{\rm CARMENES}$&Relative RV Offset of CARMENES data (m/s)&$-16831.1^{+8.3}_{-8.2}$\\
$\gamma_{\rm HARPS-N}$&Relative RV Offset HARPS-N data (m/s)&$-16289.4^{+8.7}_{-8.8}$\\
$\gamma_{\rm TRESS}$$^{a}$&Relative RV Offset of TRESS data (m/s)&$0^{+14}_{-13}$\\
$\sigma_J$&RV Jitter measured from CARMENES data (m/s)&$20.7^{+12}_{-6.5}$\\
$\sigma_J$&RV Jitter measured from HARPS-N data (m/s)&$28.1^{+9.6}_{-6.4}$\\
$\sigma_J$&RV Jitter measured from TRESS data (m/s)&$42^{+15}_{-10.}$\\
\hline

\hline
\end{tabular}
{$^a$An arbitrary number had been substracted from supplied TRESS data by Bakos {\it et al.} (2012).}
\end{table}
\clearpage
\subsection{Orbital Period Analysis}
We calculated the mid-transit times for each observation based on a reference mid-transit time (ETD number 160, observer:Yves Jongen, 2019-01-07) and an orbital period (Bakos {\it et al.} 2012) and investigated the deviations from the observed mid-transit times we derived from our models with {\sc exofast}. We then plotted these deviations with respect to epoch of observation and formed the TTV diagram that is shown in Figure-6. We fitted a linear (Model 1) and a second degree polynomial model (Model 2) independently to the data to correct the linear ephemeris and search for a potential orbital period change. We used {\sc emcee} (Foreman-Mackey {\it et al.} 2013) code for fitting procedure by making use of 500 random walkers, each of which was iterated for 5000 steps. We discarded the first 500 chains for the burn-in phase. We generated random samples for fit parameters and computed the likelihood of each sample based on its agreement with the TTV diagram. Posterior probability distribution of each of the fit parameters was computed, from which the median value and 1-$\sigma$ uncertainties of the fit parameters are obtained. As a result, we obtained posterior probability distributions for 2 parameters (slope and y-intercept) for the linear fit and 3 parameters (y-intercept, slope and quadratic term) for the polynomial fit. We added y-intercept from both models to the reference transit mid time (T$_0$), slope parameter to the orbital period to correct the reference light elements for future transit observations. 

New ephemeris information was derived from the linear model (model 1) as

\begin{equation}{
\mathcal{T} = 2458490.654297(32) + 1.327346835(12)\times E
}
    \label{eq:linear_eph}
\end{equation}

while ephemeris derived from parabolic model (model 2) as

\begin{equation}{
\mathcal{T} = 2458490.654208(35) + 1.327347225(73)\times E + 285(47)\times 10^{-10}\times  E^2
}
    \label{eq:quad_eph}
\end{equation}

Both Bayesian (BIC) and Akaike (AIC) Information Criteria strongly favour the Model 2 with $\Delta$BIC 31.2 and $\Delta$AIC 33.8. On the other hand, chi-squared and reduced chi-squared of Model 1 is $\chi^2$ = 553.64 and $\chi^2_\nu$ = 5.95 while $\chi^2$ = 517.89 and $\chi^2_\nu$ = 5.63 for Model 2 meaning parabola is slightly better than linear fit but both models are not sufficently represents the data or the error bars are underestimated. Nevertheless, Model 2 indicates an increase in the orbital period by $\dot{\rm P} = 0.14\pm0.02$ seconds in 10 years rather than a decline as expected. However, we did not find any statistically significant parabolic trend in transit times from undetrended light curves which could be lost in scatter if the trend is real.

We also applied the same procedure to a TTV diagram, constructed based on the undetrended data. But the standard deviations of residuals from the linear fit was $\sigma$ = 1.8 minutes while it is $\sigma$ = 1.6 minutes for the data detrended from red-noise with GP. The scatter around the linear model decreased significantly thanks to detrending especially for the well sampled intervals like that covered by TESS observations. 

The linear and quadratic models; however, resulted in large reduced chi-square values, hinting existence of another potential variation and/or underestimation of the error bars. Therefore, we performed frequency analyses on both data sets to investigate high-frequency variations explaining the scatter around the linear models larger than the error bars suggest. We found a cyclic variation with 15.85 day-periodicity in the TTV diagram, based on the detrended data set and corrected for the linear trend,  with a FAP of 0.06. But we did not find any significant cyclic variation in the TTV diagram constructed with the undetrended data. This cyclic variation has an amplitude compatible with the scatter of the data. Therefore we noted the frequency we found for future investigations.

The data we used to form Figure-6 is available in machine readable format in detail (i.e. including $\beta$ and PNR values, mid-transit times and their error bars and also sources of the data). 

\begin{figure}[!ht]
\begin{center}
\includegraphics[width=\textwidth]{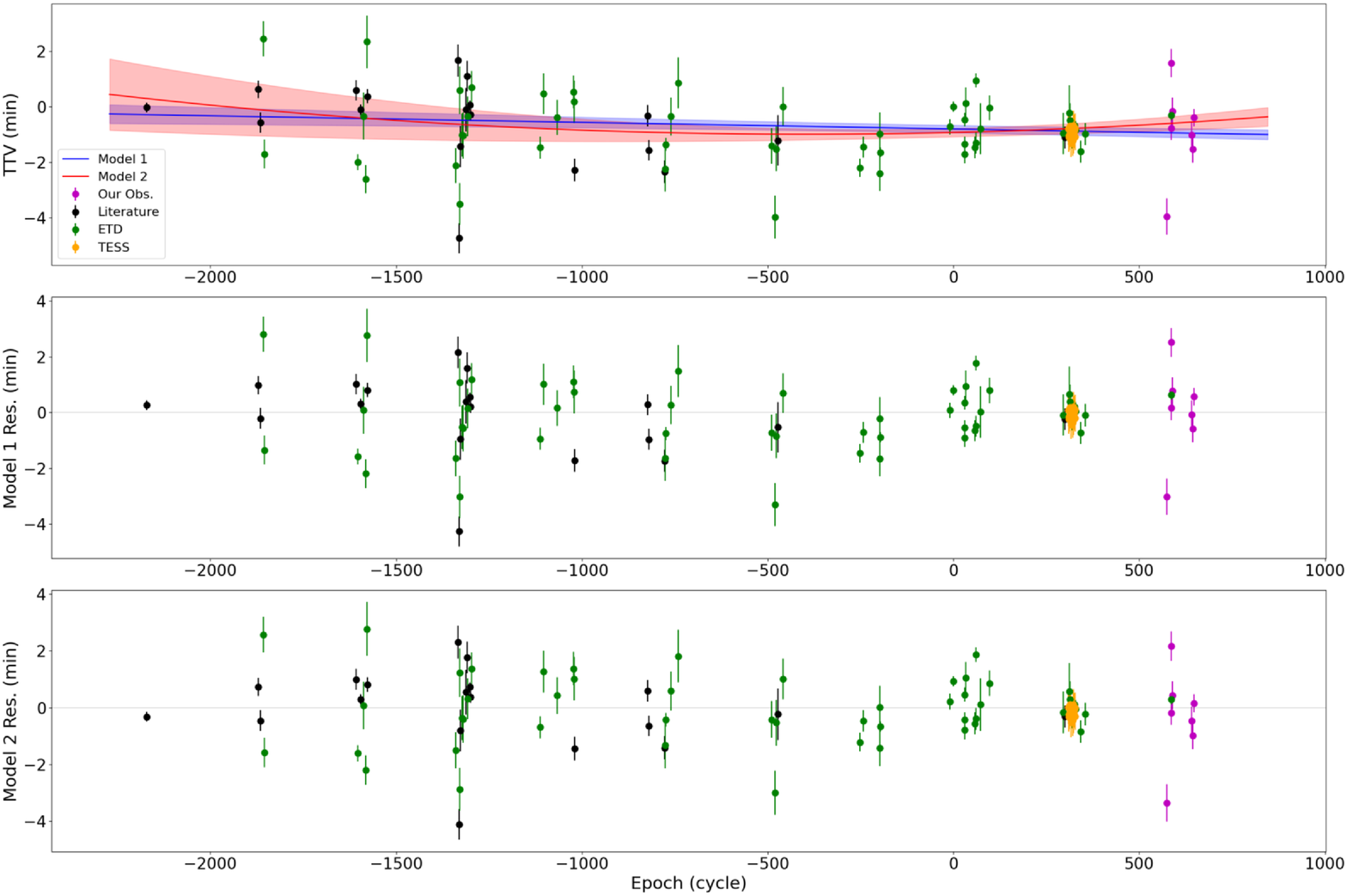}
\FigCap{Top panel: TTV of HAT-P-36 b calculated using detrended light curves. Source of data values are labeled in the legend. Red and blue lines are best parabolic and linear models respectively. Shaded areas with the same colors show 3 $\sigma$ uncertainties of the models. Residuals from the linear and parabolic models are shown in middle and bottom panel, respectively.}
\end{center}
\end{figure}

\section{DISCUSSION}

The fundamental light and radial velocity curve parameters from our analysis are in agreement with the previous studies. We also obtained the radius of the star based on SED-analysis, confirming the value calculated before with the same method by Stassun {\it et al.} (2017) within the uncertainties of measurements. However, the absolute parameters of the host star and its planet are not in 1$\sigma$ agreement with a recent study by Wang {\it et al.} (2019) even though most other parameters do agree within the same limits. Our stellar mass and radius values ($M_{\star}$ = $0.961^{+0.055}_{-0.042}$ $M_\odot$, $R_{\star}$ = $1.052^{+0.021}_{-0.018}$ $R_\odot$) are smaller than what Wang {\it et al.} (2019) found ($M_{\star}$ =$1.049^{+0.048}_{-0.046}$ $M_\odot$, $R_{\star}$ = $1.108^{+0.025}_{-0.024}$ $R_\odot$). This difference is caused by the difference between theoretical stellar models employed in the analysis in order to calculate the mass and the radius of the star which directly affects the absolute parameters of the planet. Wang {\it et al.} (2019) used the empirical relation  given by Torres {\it et al.} (2010) based on the parameter values from transit models to calculate the stellar mass and the radius in their global modelling while we used the MIST-grids instead of any empirical relations. We also provided a prior on $R_{\star}$ from the SED model which penalizes the fits in order to improve stellar parameters (Eastman {\it et al.} 2019). 

Since HAT-P-36 is an active star displaying significant variability in and out of its transit profiles, we attempted at treating these stellar spot-induced signals on the transit light curve as correlated noise and detrend the light curves for it together with other red noise sources based on Gaussian Processes. This approach improved the accuracy (standart deviation of linear residuals reduced to 1.8 minutes from 1.6 minutes) and precision (mean mid time errors are reduced to 0.52 minutes from 0.74 minutes) of the measurements of mid-transit times significantly. Therefore we constructed another TTV diagram based on these measurements from light curves detrended for the red-noise. We then compared the TTV results from these detrended and undetrended data sets and we found out that detrending reduces the scatter around the linear models significantly. We corrected the linear ephemeris (Equation-3), as a result, based on the parameters of the linear model of the TTV diagram constructed with the timings measured from the detrended light curves. Since the scatter around this linear model suggested a variation larger than the error bars of the measurements, we constructed Lomb-Scargle periodograms of the TTV and found a tentative 15.85 day-periodicity with 6\% FAP only in the TTVs based on the detrended light curves, which we noted for further observations.

Our approach based on detrending the light curves from red-noise due to spot-induced asymmetries or any other source also increased the statistical significance of the quadratic model of the TTV diagram, which turned out to be more probable statistically than the linear model. However, the reduced chi-square value of $\chi_\nu$ = 5.63 for the quadratic model is also not reassuring when the sampling and a potential underestimation of the error bars in measurements are considered.  Nevertheless, the quadratic coefficient of the best-fitting parabola is positive indicating an increase in the orbital period despite the expectancy of an orbital decay based on system parameters. Hence, we can only provide a lower limit on the tidal quality parameter for the host star as $Q^{\prime}_{\star} > (6.97 \pm 1.63) \times 10^4$ following Goldreich \& Soter (1966) and Ogilvie (2014). If the signal is real, then this parabolic trend could also be a short segment of a longer term cyclic variation.

We observed that spot-induced hump features on transit profiles in TESS (and also other) light curves pop-up and then disappear almost randomly. These behaviours and rather low amplitude of the signal ($\sim 0.5$ milimags) hint that the spot groups observed in TESS sector-22 may have short lifetimes, not cover a large area, and / or have low temperature factors. This can be related to the short observing window, which can coincide with a quiet phase of the star in its magnetic activity cycle. The predominant type of activity (i.e., spot or faculae domination) on the surface of solar-like stars is also a controlling factor and affect the detectability of stellar rotation period (Reinhold {\it et al.} 2021). Instrumental systematics and / or reduction procedure may have also introduced the variability at this frequency. Due to these ambiguities, disagreement with the projected rotational velocity from high-resolution spectroscopy (Bakos {\it et al.} 2012), data precision and observing window issues, we do not adopt the periodicity at 4.22 days we found from the out-of-transit variability in the PDCSAP fluxes of HAT-P-36 as its rotation period. However, we encourage observers to observe the star spectroscopically as well as photometrically in addition to the upcoming sector-49 observations with TESS at the end of February 2022.


\Acknow{We are grateful to Prof. Dr. Berahitdin Albayrak, whom we lost in a tragic train accident at the end of 2018, for his efforts in providing and installing T80 in AUKR campus. This work is supported by the research fund of Ankara University (BAP) through the project 18A0759001. OB gratefully acknowledges T\"UB{\.I}TAK for their support with the research grant 118F042. We thank T\"UB{\.I}TAK National Observatory of Turkey (TUG) for the partial support in using the T100 telescope with the project number 19AT100-1471. FH acknowledges the internship program provided by AUKR.}

\end{document}